\documentclass[12pt]{revtex4}
\usepackage[latin1]{inputenc}\usepackage[brazil]{babel}
\usepackage[dvips]{graphics}
\newcommand{\rem}[1]{{\bf Remark:}}

\def\QED{{\hspace*{\fill}{\vrule height .5ex width 1ex }\quad} 
    \vskip 0pt plus20pt}
\newcommand{\be}{\begin{equation}}
\newcommand{\ee}{\end{equation}}
\newcommand{\bea}{\begin{eqnarray}}
\newcommand{\eea}{\end{eqnarray}}
\newcommand{\beann}{\begin{eqnarray*}}
\newcommand{\eeann}{\end{eqnarray*}}

\begin{document}
\title{Textos de f\'{\i}sica para professores e estudantes}
\author{\bf Oscar Bolina \footnote{Agrade\c co \`a FAPESP o apoio
financeiro pelo projeto $01/08485-6$.}}
\affiliation{{\sc Departamento de F\'{\i}sica-Matem\'atica \\
Universidade de S\~ao Paulo\\
Caixa Postal 66318 \\
S\~ao Paulo 05315-970 Brasil} \\
{\bf E-mail:} {\rm bolina@if.usp.br}}
\maketitle
\begin{center}
{\bf Resumo}
\end{center}
{\sl Este texto comp\~oe-se de quatro pequenas notas -- independentes
entre si -- em que se discutem alguns t\'opicos espec\'{\i}ficos de 
f\'{\i}sica para o segundo grau e o primeiro ano universit\'ario.}
\vskip .3 cm
\noindent
\begin{center}
{\bf Abstract}
\end{center}
{\sl This is a series of short teaching papers dealing with
specific topics in a standard first-year undergraduate Physics 
course.}
\newpage
\begin{center}
{\bf \large  \'Indice}
\end{center}
\tableofcontents
\newpage
\section{M\'edia aritm\'etica, m\'edia harm\^onica e velocidade m\'edia}
\noindent
A m\'edia aritm\'etica de uma s\'erie de n\'umeros \'e a soma dos 
n\'umeros da s\'erie dividida pelo n\'umero de termos da s\'erie. 
Assim a m\'edia aritm\'etica dos n\'umeros $x_{1}=100$, $x_{2}=200$, 
$x_{3}=300$, $x_{4}=400$
\'e
\be\label{media}
\mbox{m\'edia~aritm\'etica}=\frac{x_{1}+x_{2}+x_{3}+x_{4}}{4}
=\frac{100+200+300+400}{4}=\frac{1000}{4}=250.
\ee
Qual \'e a rela\c c\~ao entre a m\'edia aritm\'etica e o conceito 
de velocidade m\'edia? Para responder a essa quest\~ao, imagine um 
avi\~ao cuja rota de v\^oo tem quatro trechos, cada um de 
$100~{\rm km}$ de comprimento \cite{M}. O avi\~ao percorre o primeiro 
trecho com velocidade de $100~{\rm km}/{\rm h}$, o segundo com 
velocidade de $200~{\rm km}/{\rm h}$, o terceiro com velocidade de 
$300~{\rm km}/{\rm h}$, e o quarto trecho com velocidade de 
$400~{\rm km}/{\rm h}$. Qual \'e a velocidade m\'edia do avi\~ao em 
sua rota? Se usarmos a m\'edia aritm\'etica, encontraremos
\be\label{ma}
\mbox{Velocidade~m\'edia}=\frac{100~{\rm km}/{\rm h}+200~{\rm km}/{\rm
h}+300~{\rm km}/{\rm h}+400~{\rm km}/{\rm h}}{4}=250~{\rm km}/{\rm h},
\ee
como em (\ref{media}). A m\'edia aritm\'etica, no entanto, d\'a um
resultado errado para a velocidade m\'edia do avi\~ao. Isso porque
as velocidades do avi\~ao em cada trecho do percurso, apesar de
serem mantidas pela mesma dist\^ancia, n\~ao s\~ao mantidas pelo
mesmo tempo de v\^oo. Al\'em disso, a velocidade m\'edia \'e sempre
a raz\~ao da dist\^ancia percorrida pelo tempo gasto para percorrer a 
dist\^ancia, e essa id\'eia n\~ao foi usada em (\ref{ma}). A velocidade 
m\'edia correta do avi\~ao \'e calculada da seguinte maneira.
\[
\begin{array}{ccccc}
{\rm Tempo~para~percorrer~o~primeiro~trecho} ~&~1~{\rm h} \\
{\rm Tempo~para~percorrer~o~segundo~trecho} ~&~30~{\rm minutos} \\
{\rm Tempo~para~percorrer~o~terceiro~trecho} ~&~20~{\rm minutos} \\
{\rm Tempo~para~percorrer~~o~~quarto~trecho} ~&~15~{\rm minutos} \\
---------------- & ------ \\
{\rm ~~Tempo~~total~~de~~percurso~~} ~&~2~{\rm h}~{\rm e}~5~{\rm minutos}
\\
\end{array}
\]
O avi\~ao ent\~ao percorre a dist\^ancia total de $400~{\rm km}$ em
$2~{\rm h}~5~{\rm minutos}$. A velocidade m\'edia calculada
pela defini\c c\~ao ser\'a assim
\be\label{medh}
\mbox{Velocidade~m\'edia}=\frac{400~{\rm km}}{\frac{25}{12}~{\rm
h}}=192~{\rm km}/{\rm h}.
\ee
\`A m\'edia que acabamos de calcular d\'a-se o nome de {\sl m\'edia
harm\^onica}. A m\'edia harm\^onica \'e definida como o inverso da
m\'edia aritm\'etica do inverso dos n\'umeros para os quais se
deseja calcular a m\'edia:
\be\label{mh}
\mbox{M\'edia~harm\^onica}=
\frac{1}{\left (\frac{\frac{1}{x_{1}}+\frac{1}{x_{2}} + \cdot \cdot \cdot 
+\frac{1}{x_{n}}}{n} \right )}=\frac{n}{\frac{1}{x_{1}}+\frac{1}{x_{2}}+ 
\cdot \cdot \cdot +\frac{1}{x_{n}}}.
\ee
Essa \'e a m\'edia a ser usada em problemas que envolvem raz\~ao de
varia\c c\~ao, como velocidades (quil\^ometros por hora) e pre\c cos
(reais por d\'uzia). Vamos usar (\ref{mh}) no problema do 
avi\~ao mencionado acima. Temos
\[
\mbox{M\'edia~harm\^onica~da~velocidade}=
\frac{4}{\frac{1}{100\frac{{\rm km}}{\rm h}}+\frac{1}{200\frac{{\rm
km}}{\rm h}}+\frac{1}{300\frac{{\rm km}}{\rm h}}
+\frac{1}{400\frac{{\rm km}}{\rm h}}}
=\frac{4 \times 1200}{25}~{\rm km}/{\rm h}=192~{\rm km}/{\rm h},
\]
como em (\ref{medh}).
\subsection{Exemplos}
\noindent
{\bf 1)} Kelly Quina vai e volta de carro de Santos a Bertioga em 
busca de seu cachorrinho. Seu carro faz 16 quil\^ometros por litro 
de gasolina na viagem de ida, e 12 quil\^ometros por litro na viagem
de volta. Calcule a m\'edia harm\^onica do consumo de gasolina em   
quil\^ometros por litro. Se a dist\^ancia de Santos a Bertioga \'e de
$60~{\rm km}$, verifique que a m\'edia harm\^onica \'e a m\'edia
correta a ser calculada. Encontre a m\'edia aritm\'etica para
comparar. Discuta seus resultados.
\vskip .5 cm
\noindent
{\bf 2)} O pre\c co (em reais) da d\'uzia de ovos por semana em
certo m\^es \'e dado na tabela abaixo:  
\[
\begin{array}{ccccc}
\mbox{semana}~&~|~&~\mbox{pre\c co} \\ 
----~&~|~&----\\
1~&~|~&~{\rm R} \$~ 2,15 \\
2~&~|~&~{\rm R} \$~ 2,48 \\
3~&~|~&~{\rm R} \$~ 2,36 \\
4~&~|~&~{\rm R} \$~ 2,07
\end{array}
\]
Calcule a m\'edia aritm\'etica e a m\'edia harm\^onica do pre\c co
da d\'uzia de ovos no m\^es em quest\~ao. Qual \'e o significado de 
cada uma delas neste exemplo?
\vskip .5 cm
\noindent
{\bf 3)} Um motorista percorre uma estrada de comprimento $d$ 
mantendo uma velocidade constante $v_{1}$ por um trecho de
comprimento $x$ ($x < d$). Que velocidade constante deve o 
motorista manter no trecho restante da estrada para que sua 
velocidade m\'edia no percurso total seja a m\'edia aritm\'etica 
das velocidades nos dois trechos?
\vskip .5 cm
\noindent
{\bf 4)} Voc\^e dirige seu carro por 1 quil\^ometro subindo a
serra com velocidade de $25~{\rm km}/{\rm h}$. Que velocidade 
voc\^e deve manter no quil\^ometro de descida do outro lado da 
serra para que sua velocidade m\'edia no percurso total de 
2 quil\^ometros seja de $50~{\rm km}/{\rm h}$? (Veja esse e
outros problemas endiabrados em \cite{N}.)
\vskip .5 cm
\noindent
{\bf 5)} Demonstre que se um carro mant\'em uma velocidade m\'edia
$v_{1}$ por certo intervalo de tempo, e em seguida mant\'em uma 
velocidade m\'edia $v_{2}$ pelo mesmo intervalo de tempo, 
ent\~ao a velocidade m\'edia do carro no intervalo de tempo
total de percurso \'e dada pela m\'edia aritm\'etica
\[
v_{m}=\frac{v_{1}+v_{2}}{2}.
\]
\vskip .5 cm
\noindent
\begin{center}
{\bf Refer\^encias}
\end{center}

\vfill \hrule width2truein \smallskip {\baselineskip=10pt \noindent 
{\small S\~ao doze horas. A que horas voc\^e pretende 
terminar esse problema? -- Logo que os dois ponteiros estiverem 
separados e em linha reta! A que horas estar\'a terminado o problema?
(Mello e Souza, {\sl Diabruras da Matem\'atica},
Editora Get\'ulio Costa, 1943, p.187.)}
\par 
}

\section{O movimento uniformemente acelerado}
\begin{figure}
\begin{center}
\resizebox{!}{1.5 truecm}{\includegraphics{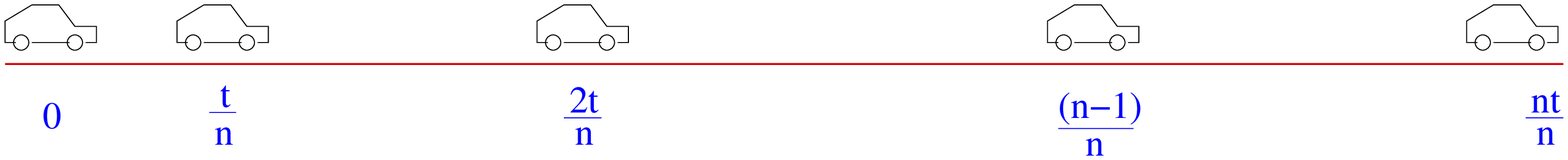}}
\vskip .2 cm
\parbox{13truecm}{\caption{\baselineskip=16 pt\small\label{str}
\noindent
A aproxima\c c\~ao de um movimento uniformemente acelerado por uma 
sequ\^encia de movimentos uniformes em intervalos de tempo de mesma 
dura\c c\~ao. Como a velocidade do corpo varia ao cabo de cada 
subintervalo, os sucessivos deslocamentos do corpo aumentam ou diminuem 
conforme a acelera\c c\~ao \'e positiva ou negativa.}}
\end{center}
\end{figure}

\noindent
O movimento acelerado mais simples \'e o movimento unidimensional 
com acelera\c c\~ao constante. As equa\c c\~oes que descrevem esse 
movimento s\~ao bem conhecidas. O deslocamento de um corpo que se move 
com acelera\c c\~ao constante $a$ \'e uma fun\c c\~ao 
do tempo $t$ dada por
\be\label{um}
s=v_{0}t+\frac{1}{2}at^{2},
\ee
sendo que $v_{0}$ \'e a velocidade inicial do corpo. H\'a uma 
maneira de obter a equa\c c\~ao (\ref{um}) a partir
da aproxima\c c\~ao do movimento acelerado por uma sequ\^encia 
de movimentos uniformes em intervalos de tempo muito curtos. 
\newline
O processo se assemelha ao problema do c\'alculo de encontrar 
a \'area sob uma curva dada pela soma de \'areas de ret\^angulos,
sendo essa soma uma aproxima\c c\~ao para a \'area dada quando os 
ret\^angulos se tornam cada vez menores.
\newline
Seja um corpo que se move com velocidade inicial $v_{0}$ e
acelera\c c\~ao constante $a$, e suponha ser $s$ o deslocamento 
do corpo em certo intervalo de tempo $t$. Esse movimento uniformemente
acelerado pode ser aproximado por movimentos uniformes da seguinte 
maneira. Imagine que o intervalo de tempo $t$ seja dividido em
certo n\'umero inteiro $n$ de subintervalos de tempo de dura\c c\~ao 
igual a $\frac{t}{n}$, e que em cada um desses subintervalos 
a  velocidade do corpo seja constante. A situa\c c\~ao \'e ilustrada na 
{\sc figura} \ref{str}. Assim, o movimento acelerado \'e aproximado por 
$n$ movimentos uniformes nos seguintes subintervalos de tempo:
\[
0 \leftrightarrow \frac{t}{n}, \;\;\;\;\;\; 
\frac{t}{n} \leftrightarrow \frac{2t}{n}, \;\;\;\;\;\; 
\frac{2t}{n} \leftrightarrow \frac{3t}{n},
\;\;\;\;\;\; . \;\;\;\; . \;\;\;\;  ., \;\;\;\;\;\; \frac{(n-1)t}{n} 
\leftrightarrow 
\frac{nt}{n}.
\]
Para imitar um movimento acelerado, no entanto, vamos supor que a 
velocidade do corpo varie instantaneamente no final de cada 
subintervalo de tempo. Como a acelera\c c\~ao do movimento que
se quer aproximar \'e constante, vamos impor que haja um 
incremento constante de velocidade igual a $\frac{at}{n}$ ao 
cabo de cada subintervalo de tempo. Assim, as velocidades do 
corpo em cada subintervalo de tempo ser\~ao iguais a
\[
v_{0}, \;\;\; v_{0}+\frac{at}{n}, \;\;\; v_{0}+\frac{2at}{n}, \;\;\;
. \;\; . \;\; . \;\;\; v_{0}+\frac{(n-1)at}{n}.
\]
Como a sequ\^encia de movimentos uniformes \'e para ser uma aproxima\c 
c\~ao ao movimento acelerado do corpo, o deslocamento $s$ no movimento
acelerado deve ser aproximado pela soma dos $n$ deslocamentos do corpo
em cada um dos $n$ subintervalos de tempo. Seja $s_{n}$ essa soma. 
Sendo a velocidade constante em cada subintervalo, $s_{n}$ 
\'e simplesmente o produto da velocidade do corpo pela dura\c c\~ao 
do subintervalo de tempo. Disso vem que $s_{n}$ \'e
\[
s_{n}=v_{0}\cdot \frac{t}{n} +\left (v_{0}+\frac{at}{n} 
\right )\frac{t}{n} + 
\left (v_{0}+\frac{2at}{n} \right )\frac{t}{n} + . . . +  
\left (v_{0}+\frac{(n-1)at}{n} \right )\frac{t}{n}.
\]
A simplifica\c c\~ao da express\~ao acima nos leva a 
\be\label{2}
s_{n}=v_{0}t + \frac{at^{2}}{n^{2}} \left (1~+~2~+~3~+~.~ .~ .~ +~n-1 
\right ). 
\ee
O termo entre par\^enteses em (\ref{2}) \'e a soma aritm\'etica dos
$n-1$ primeiros n\'umeros inteiros. A f\'ormula para essa soma \'e
\[
1~+~2~+~3~+~.~ .~ .~ +~n-1=\frac{n(n-1)}{2}.
\]
Portanto, $s_{n}$ fica
\be\label{3}
s_{n}=v_{0}t + \frac{at^{2}}{n^{2}} \cdot \frac{n(n-1)}{2}
=v_{0}t+\frac{1}{2}at^{2} \left (1-\frac{1}{n^{2}} \right ).
\ee
Como est\~ao $s$ e $s_{n}$ relacionados? Se fizermos $n$ cada 
vez maior, o movimento acelerado ser\'a aproximado por um 
n\'umero cada vez maior de movimentos uniformes, cada um deles 
ocorrendo em subintervalos de tempo cada vez menores (iguais a
$t/n$). Assim, a varia\c c\~ao da velocidade que ocorre entre 
subintervalos vai se aproximar cada vez mais de uma varia\c c\~ao 
cont\'{i}nua de velocidade. \'E no caso extremo em que o n\'umero 
$n$ de subintervalos \'e infinito que a aproxima\c c\~ao por movimentos 
uniformes \'e o pr\'oprio movimento acelerado, e nesse caso $s_{n}$ \'e 
$s$. Mas quando $n$ \'e infinito, o termo $1/n^{2}$ entre par\^enteses em 
(\ref{3}) \'e zero, e $s_{n}$ passa a ser
\be\label{4}
s=v_{0}t+\frac{1}{2}at^{2},
\ee
que \'e a deriva\c c\~ao alternativa de (\ref{um}). No c\'alculo, a 
passagem de (\ref{3}) a (\ref{4}) \'e sintetizada ao ser dizer que {\sl o 
limite de} $s_{n}$ {\sl quando} $n$ {\sl tende a infinito \'e $s$}, e a 
nota\c c\~ao para isso \'e
\[
\lim_{n \longrightarrow \infty} s_{n}=s.
\]
\subsection{Exemplos}
\noindent
{\bf 1)} O caso mais simples de movimento com acelera\c c\~ao vari\'avel 
\'e aquele em que a acelera\c c\~ao varia linearmente com o tempo. 
Aproxime tal movimento por uma sequ\^encia de movimentos com acelera\c 
c\~ao constante em subintervalos de tempo de dura\c c\~ao $\frac{t}{n}$. 
Suponha que a acelera\c c\~ao varie instantaneamente somente ao cabo de 
cada subintervalo de tempo por incrementos iguais a $\frac{bt}{n}$, sendo 
que a constante $b$ neste caso \'e a raz\~ao de varia\c c\~ao da acelera\c 
c\~ao.
\newline
Se o corpo tem velocidade inicial $v_{0}$ e acelera\c c\~ao 
inicial $a_{0}$, use a f\'ormula (\ref{4}) para calcular o 
deslocamento do corpo em cada subintervalo de tempo, some esses 
deslocamentos e tome o limite $n \rightarrow \infty$ para mostrar 
que o deslocamento neste movimento com acelera\c c\~ao vari\'avel 
\'e dado por
\[
s=v_{0}t + \frac{a_{0}}{2}t^{2} +\frac{b}{6}t^{3}. 
\]
{\bf Dica:} {\sl Muito provavelmente voc\^e precisar\'a da f\'ormula
\[
1^{2}~+~2^{2}~+~3^{2}~+~\cdot \cdot \cdot ~+~n^{2}=
\frac{n}{6}+\frac{n^{2}}{2}+\frac{n^{3}}{3}.
\]
}
\vfill
\hrule width2truein \smallskip {\baselineskip=10pt \noindent
{\small Jo\~ao e Carlos t\^em apenas uma bicicleta e devem fazer um 
percurso de $32~{\rm km}$. Jo\~ao partir\'a de bicicleta e deix\'a-la-\'a 
num certo ponto combinado do caminho, continuando a viagem a p\'e. Carlos 
percorrer\'a o primeiro trecho a p\'e, retomar\'a a bicicleta deixada no 
ponto combinado e concluir\'a o percurso na bicicleta; se os seus 
c\'alculos estiverem certos, chegar\~ao simultaneamente ao fim do 
percurso. Se ambos, de bicicleta, fazem $30~{\rm km}$ por hora, Jo\~ao
fazendo, a p\'e, $5~{\rm km}$ por hora e Carlos $6~{\rm km}$ por hora, 
pergunta-se: a que dist\^ancia do ponto de partida deve Jo\~ao deixar a 
bicicleta? (Mello e Souza, {\sl Diabruras da Matem\'atica},
Editora Get\'ulio Costa, 1943, p.176.)}
\par
}

\newpage
\section{A par\'abola e o movimento parab\'olico}
\begin{figure}
\begin{center}
\resizebox{!}{5.8 truecm}{\includegraphics{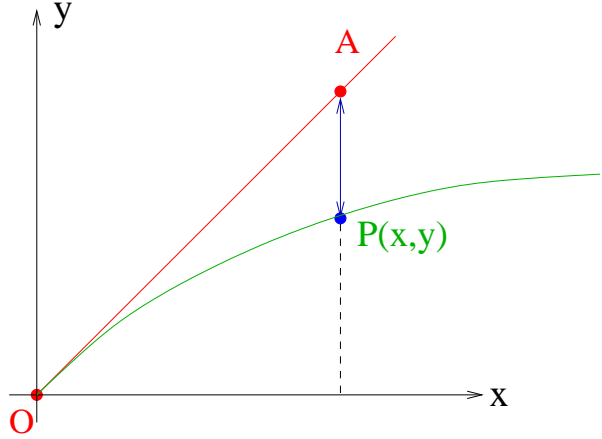}}
\vskip .2 cm
\parbox{13truecm}{\caption{\baselineskip=16 pt\small\label{parabola}
\noindent
Deriva\c c\~ao da trajet\'oria parab\'olica de um proj\'etil pela 
composi\c c\~ao de dois movimentos, um ao longo da reta $OA$ com 
velocidade constante $v_{0}$, outro de queda livre sob a a\c c\~ao da 
gravidade. A posi\c c\~ao final do proj\'etil \'e o ponto $P$ na 
par\'abola.}
}
\end{center}
\end{figure}

\begin{figure}
\begin{center}
\resizebox{!}{5 truecm}{\includegraphics{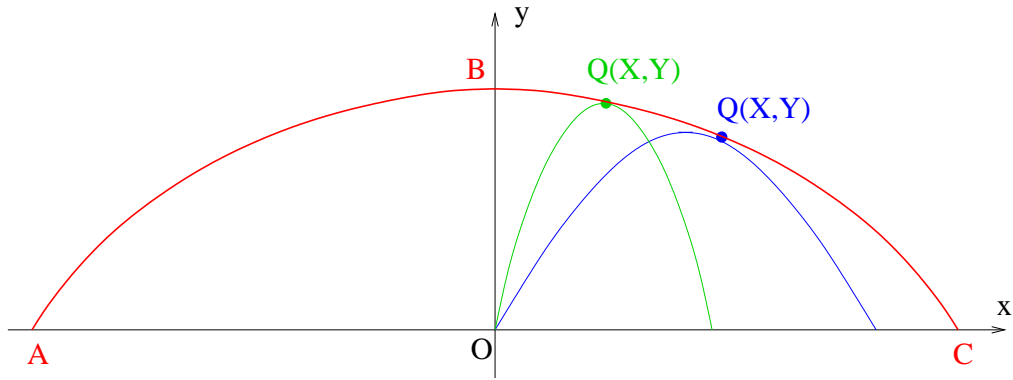}}
\vskip .1 cm
\parbox{13truecm}{\caption{\baselineskip=16 pt\small\label{parabola5}
\noindent
A par\'abola de seguran\c ca $ABC$ tem um e somente um ponto em 
comum com as par\'abolas descritas por proj\'eteis disparados de 
$O$, com a mesma velocidade inicial $v_{0}$ e diferentes \^angulos 
de lan\c camentos, para as quais $\Delta =0$.
}}
\end{center}
\end{figure}

\noindent
Um proj\'etil \'e disparado de um ponto $O$ com velocidade inicial 
$v_{0}$ em uma dire\c c\~ao que forma um \^angulo de lan\c camento
$\theta$ com a horizontal como mostrado na {\sc figura} 
\ref{parabola}. 
\newline
Se a for\c ca da gravidade n\~ao existisse, o movimento do
proj\'etil seria ao longo da linha reta $OA$ com velocidade 
constante $v_{0}$. Assim, depois de $t$ segundos, o proj\'etil
estaria em $A$, a uma dist\^ancia 
\[
OA=v_{0}t
\]
do ponto $O$.
\newline
A for\c ca da gravidade, no entanto, agindo sobre o proj\'etil 
em $t$ segundos, imprime a ele uma acelera\c c\~ao constante 
$g$, o que o traz a um ponto $P$ situado a uma dist\^ancia  
\[
AP=\frac{1}{2} gt^{2}
\]
abaixo do ponto $O$. Os movimentos do proj\'etil ao longo de $OA$ e 
$AP$ s\~ao independentes, e a composi\c c\~ao deles dar\'a a
posi\c c\~ao $P$ do proj\'etil ao cabo de $t$ segundos. Se as 
coordenadas do ponto $P$ em rela\c c\~ao ao sistema de eixos 
retangulares definido no plano do movimento como
indicado na {\sc figura} \ref{parabola} s\~ao $(x,y)$, 
ent\~ao 
\be\label{1111}
x=OA \cdot \cos{\theta}=v_{0}t\cos\theta 
\ee
e
\be\label{2222}
y=OA\sin{\theta}-AP=v_{0}t\sin{\theta}-\frac{1}{2}gt^{2}.
\ee
Ao se eliminar $t$ entre as equa\c c\~oes (\ref{1111}) e (\ref{2222}),
obt\'em-se uma equa\c c\~ao para a coordenada $y$ como fun\c c\~ao
da coordenada $x$, a qual representa a equa\c c\~ao da trajet\'oria 
do proj\'etil disparado do ponto $O$ com velocidade inicial de lan\c 
camento $v_{0}$ e \^angulo de lan\c camento $\theta$. Tem-se
\be\label{3333}
y=v_{0} \left (\frac{x}{v_{0}\cos{\theta}}\right )\sin{\theta}
-\frac{1}{2}g \left ( \frac{x}{v_{0}\cos{\theta}} \right )^{2}
=x\tan{\theta}-\frac{g x^{2}}{2 v_{0}^{2}\cos^{2}\theta}.
\ee
Essa equa\c c\~ao quadr\'atica em $x$ representa uma par\'abola,
no plano do movimento do proj\'etil, com concavidade para baixo 
devido ao sinal negativo do coeficiente de $x^{2}$.
\newline
\noindent
Seja agora um proj\'etil disparado de $O$ com velocidade inicial de 
lan\c camento $v_{0}$. Qual deve ser o \^angulo de lan\c camento
$\theta$ para que o proj\'etil passe por certo ponto dado $Q$ de 
coordenada $(X,Y)$? A {\sc figura} \ref{parabola5} mostra duas
par\'abolas que passam por dois pontos distintos no plano. 
Como o ponto 
$Q$ est\'a em uma par\'abola, suas coordenadas devem satisfazer a 
equa\c c\~ao (\ref{3333}). Disso vem que
\be\label{yx}
Y=X\tan{\theta}-\frac{g X^{2}}{2 v_{0}^{2}\cos^{2}\theta},
\ee
e esta \'e agora uma equa\c c\~ao para o \^angulo $\theta$ de lan\c 
camento do proj\'etil. A identidade trigonom\'etrica
\[
\cos^{2}{\theta}=\frac{1}{1+\tan^{2}{\theta}},
\]
quando substitu\'{i}da em (\ref{yx}), fornece a seguinte 
equa\c c\~ao do segundo grau para a tangente do
\^angulo de lan\c camento:
\be\label{gX}
gX^{2}\tan^{2}{\theta}-2Xv^{2}_{0}\tan{\theta}
+\left (2v^{2}_{0}Y+gX^{2} \right )=0.
\ee
Como uma equa\c c\~ao do segundo grau para $\tan{\theta}$, 
a equa\c c\~ao (\ref{gX}) ter\'a duas ra\'{\i}zes reais, uma raiz real, 
ou nenhuma raiz real quando o discriminante
\be\label{dis}
\Delta=4X^{2}v^{4}_{0}-4gX^{2}\left (2v^{2}_{0}Y+gX^{2} \right )
\ee
for, respectivamente, positivo, zero ou negativo. As condi\c c\~oes
no discriminante levam \`a seguinte interpreta\c c\~ao
f\'{\i}sica do problema \cite{SL}.
\begin{itemize}
\item[{\bf 1.}] Quando $\Delta < 0$, a equa\c c\~ao (\ref{gX}) n\~ao 
tem ra\'{\i}zes reais. Portanto, nenhuma par\'abola passar\'a pelo ponto 
dado, n\~ao importa qual seja o \^angulo de lan\c camento do proj\'etil.
\vskip .2 cm
\noindent 
\item[{\bf 2.}] Quando $\Delta >0$, a equa\c c\~ao 
(\ref{gX}) tem duas ra\'{i}zes reais. Portanto, duas 
par\'abolas v\~ao passar por $Q$, correspondendo a dois 
\^angulos distintos de lan\c camento do proj\'etil. 
\vskip .2 cm
\noindent 
\item[{\bf 3.}] Quando $\Delta=0$, a equa\c c\~ao (\ref{gX}) tem 
somente uma solu\c c\~ao real. Nesse caso, somente uma par\'abola 
passar\'a por $Q$. As coordenadas do ponto $Q$ s\~ao ent\~ao 
determinadas por $\Delta=0$, e disso resulta que elas s\~ao
tais que 
\be\label{dis2}
Y=\frac{v^{2}_{0}}{2g}-\frac{g}{2v^{2}_{0}}X^{2}.
\ee
Ao se variarem as cordenadas $X$ e $Y$, a equa\c c\~ao (\ref{dis2}) 
definir\'a uma curva -- ela pr\'opria uma par\'abola ! -- de pontos 
que podem ser atingidos pelos proj\'eteis disparados de $O$ com 
velocidade inicial $v_{0}$ e diferentes \^angulos de lan\c camento. 
Essa curva ter\'a um e somente um ponto em comum com cada uma das 
trajet\'orias parab\'olicas descritas pelos proj\'eteis para as quais 
$\Delta=0$. O conjunto de todos esses pontos comuns que satisfazem a
equa\c c\~ao (\ref{dis2}) \'e a par\'abola $ABC$ mostrada 
na {\sc figura} \ref{parabola5}, para a qual
\[
OA=OC=2OB=\frac{v^{2}_{0}}{g}.
\]
\end{itemize}
\noindent
\`A par\'abola $ABC$ d\'a-se o nome de {\sc par\'abola de seguran\c ca},
e a raz\~ao para isso \'e que se $Q$ for exterior \`a par\'abola de 
seguran\c ca, nenhum proj\'etil disparado de $O$ com velocidade inicial 
$v_{0}$ poder\'a atingi-lo, enquanto que se $Q$ estiver na par\'abola 
de seguran\c ca ou for interior a ela, haver\'a proj\'eteis que 
certamente o atingir\~ao.
\subsection{Exemplos}
\begin{figure}
\begin{center}
\resizebox{!}{5.5 truecm}{\includegraphics{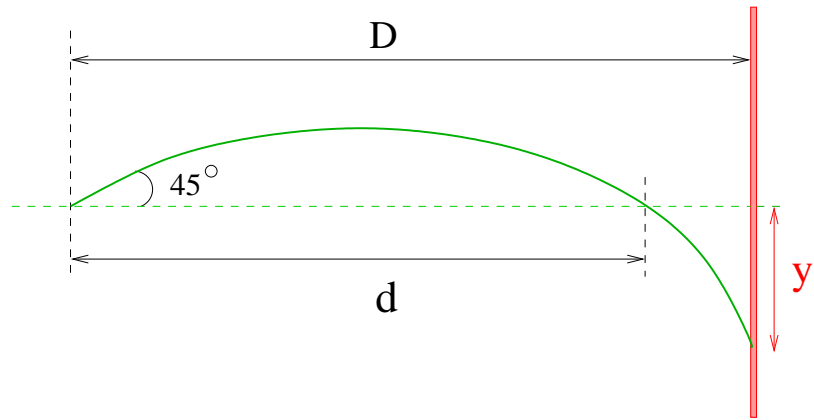}}
\vskip .2 cm
\parbox{13truecm}{\caption{\baselineskip=16 pt\small\label{parapet}
Figura referente ao problema n\'umero {\bf 12}.}
}
\end{center}
\end{figure}

\noindent
{\bf 1)} Mostre que a equa\c c\~ao da par\'abola (\ref{3333}) pode ser 
escrita na form
\[
y=-Cx(x-R)
\]
para certas constantes $C$ e $R$. Determine essas constantes e
diga qual \'e o significado f\'{\i}sico de $R$.
\vskip .8 cm
\noindent
{\bf 2)} Mostre que a equa\c c\~ao da par\'abola (\ref{3333}) pode ser 
escrita na form
\[
(x-A)^{2}=k(y-B)
\]
para certas constantes $A$, $B$ e $k$. Determine essas constantes e 
diga qual \'e o significado f\'{\i}sico de $A$ e $B$. 
\vskip .8 cm
\noindent
{\bf 3)} Mostre que se o alcance de um proj\'etil \'e $R$ e sua altura 
m\'axima \'e $H$, ent\~ao a velocidade de lan\c camento do 
proj\'etil \'e dada por
\[
\sqrt{2g \left (H+\frac{R^{2}}{16H} \right )}.
\]
\vskip .4 cm
\noindent
{\bf 4)} Uma fonte lan\c ca jatos de \'agua horizontalmente 
em todas as dire\c c\~oes de uma altura $h$ e com a mesma 
velocidade inicial $v$. Qual \'e a forma da superf\'{\i}cie 
l\'{\i}quida? Mostre que a equa\c c\~ao de uma se\c c\~ao da 
superf\'{\i}cie l\'{\i}quida cortada por uma plano horizontal 
$Oxy$ \'e dada por
\[
x^{2}+y^{2}=\frac{2hv^{2}}{g},
\]
sendo $g$ a acelera\c c\~ao da gravidade.
\vskip .8 cm
\noindent
{\bf 5)} Mostre que o alcance $R$ de um proj\'etil est\'a relacionado 
com o tempo $T$ de v\^oo do proj\'etil pela equa\c c\~ao
\[
gT^{2}=2R\tan{\theta}.
\]
\vskip .4 cm
\noindent
{\bf 6)} Um proj\'etil \'e disparado com velocidade inicial $v$ contra 
uma parede vertical situada a uma dist\^ancia $d$ do ponto de lan\c 
camento do proj\'etil. Mostre que a altura m\'axima na qual o proj\'etil
se choca contra a parede \'e dada por
\[
\frac{v^{4}-g^{2}d^{2}}{2gv^{2}}.
\]
\vskip .4 cm
\noindent
{\bf 7)} Mostre que o produto dos tempos de v\^oo que
um proj\'etil gasta para atingir dois pontos $P$ e $Q$ em 
sua trajet\'oria parab\'olica \'e dado por
\[
\frac{2 PQ}{g},
\]
sendo $PQ$ a dist\^ancia entre os pontos $P$ e $Q$.
\vskip .6 cm
\noindent
{\bf 8)} Mostre que, para a par\'abola de seguran\c ca mostrada na {\sc 
figura} \ref{parabola5}, a dist\^ancia $OB$ \'e igual \`a altura m\'axima 
alcan\c cada por um proj\'etil disparado verticalmente para cima com 
velocidade inicial $v_{0}$.
\vskip .6 cm
\noindent
{\bf 9)} Mostre que um proj\'etil disparado com velocidade inicial $v_{0}$
poder\'a atingir um ponto de coordenadas $(X,Y)$ no plano de 
sua trajet\'oria contanto que
\[
Y+\sqrt{X^{2}+Y^{2}} \leq \frac{v^{2}_{0}}{g},
\]
sendo $g$ a acelera\c c\~ao da gravidade.
\vskip .6 cm
\noindent
{\bf 10)} Um proj\'etil \'e disparado com velocidade inicial $v_{0}$
formando um \^angulo $\theta$ com a horizontal. No ponto de altura 
m\'axima de sua trajet\'oria, a velocidade do proj\'etil \'e puramente 
horizontal e de magnitude $v_{0}\cos{\theta}$. Imagine por um instante que
o proj\'etil, no ponto de altura m\'axima, est\'a se movendo em um 
c\'{\i}rculo de raio $\rho$. Use a f\'ormula da acelera\c c\~ao 
centr\'{\i}peta $a_{c}=v^{2}/\rho$, para obter o {\sl raio de curvatura 
da par\'abola} no ponto de altura m\'axima. Observe que a acelera\c c\~ao 
centr\'{\i}peta, neste caso, \'e a pr\'opria acelera\c c\~ao da gravidade. 
Discuta seu resultado.
\vskip .6 cm
\noindent
{\bf 11)} Um canh\~ao montado em um eleva\c c\~ao a uma altura $h$
do solo horizontal dispara um proj\'etil com velocidade $v_{0}$ com  
um \^angulo $\theta$ com a horizontal. Mostre que para que o alcance 
do proj\'etil no solo seja m\'aximo o \^angulo de lan\c camento
deve ser tal que 
\[
\tan{\theta}=\frac{v_{0}}{\sqrt{v^{2}_{0}+2gh}}.
\]
\vskip .3 cm
\noindent
{\bf 12)} Determine o maior valor do \^angulo de lan\c camento de 
um proj\'etil para que a dist\^ancia do proj\'etil ao ponto de 
lan\c camento seja sempre crescente.
\vskip .5 cm
\noindent
{\bf 13)} Em lan\c camentos obl\'{\i}quos de $45^{\circ}$ contra um
anteparo vertical colocado \`a dist\^ancia $D$ do ponto de lan\c camento, 
como mostrado na {\sc Figura} \ref{parapet}, mostre que a dist\^ancia 
vertical {\it y} em que o proj\'etil atinge o anteparo em fun\c c\~ao do 
par\^ametro $d$ $(d < D)$ \'e dada por
\noindent
\[
y=\frac{D^{2}}{d}-D.
\]
\vskip .5 cm
\noindent
\begin{center}
{\bf Refer\^encias}
\end{center}

\vfill
\hrule width2truein \smallskip {\baselineskip=10pt \noindent 
{\small Trecho de um romance: "Quando ela, ao sair da missa,
olhou para o rel\'ogio, em vez de dizer s\~ao dez e pouco (como
faria, no caso, sua irm\~a mais velha) exclamou risonha, tocando-me 
de leve no bra\c co:
\newline
-- V\^e, querido! Os ponteiros do rel\'ogio da Igreja j\'a formam
um \^angulo reto!
\newline
Respondi-lhe com transbordante meiguice:
\newline
-- \'Es sempre original, meu bem! Eu j\'a sabia que eram precisamente ...
"
\newline
Complete a frase final do trecho acima. (Mello e Souza, {\sl Diabruras da 
Matem\'atica},
Editora Get\'ulio Costa, 1943, p.187.)}
\par 
}

\newpage

\section{Qual \'e o plano do movimento circular?}
\begin{figure}
\begin{center}
\resizebox{!}{6.5 truecm}{\includegraphics{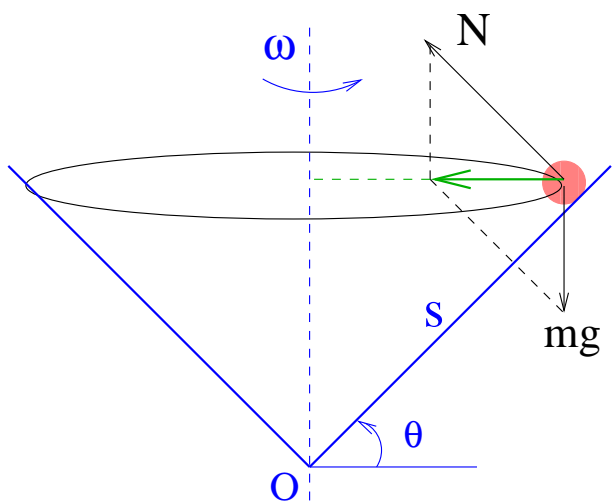}}
\vskip .1 cm
\parbox{14truecm}{\caption{\baselineskip=16 pt\small\label{conn}
\noindent
A figura illustra um problema t\'{\i}pico de movimento circular.
A esfera descreve um movimento circular de raio $s\cos{\theta}$
a uma dist\^ancia $s$ do v\'ertice do cone. A resultante da 
for\c ca peso e da rea\c c\~ao normal situa-se no plano
horizontal.}
}
\end{center}
\end{figure}

\begin{figure}
\begin{center}
\resizebox{!}{7 truecm}{\includegraphics{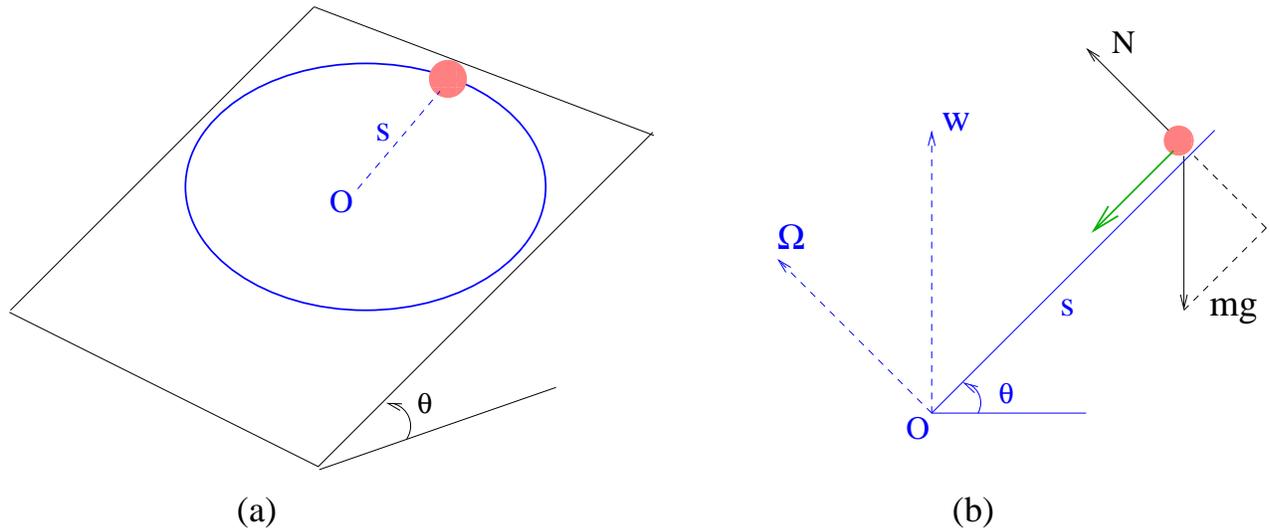}}
\vskip .1 cm
\parbox{14truecm}{\caption{\baselineskip=16 pt\small\label{con2}
Movimento circular em um plano inclinado. Sendo um vetor,
a velocidade angular do movimento se decomp\~oe como qualquer
grandeza vetorial. Disso resulta que h\'a escolhas para o plano do 
movimento 
circular.} }
\end{center}
\end{figure}

\noindent
A {\sc figura} \ref{conn} illustra um problema t\'{\i}pico do 
movimento circular. Uma esfera de massa $m$ \'e lan\c cada 
com velocidade horizontal $v$ sobre a superf\'{\i}cie interna 
de um cone de \^angulo de abertura $\frac{\pi}{2}-\theta$.
Pergunta-se qual deve ser o valor da velocidade $v$ que se deve 
imprimir \`a esfera para que ela descreva um movimento circular 
no plano horizontal do seu vetor velocidade, a uma dist\^ancia 
$s$ do v\'ertice $O$ do cone. 
\newline
Vamos aqui discutir as hip\'oteses que s\~ao rotinamente feitas para 
responder a essa quest\~ao, e contrast\'a-la com uma an\'alise 
diferente de uma variante desse problema. 
\newline
Duas for\c cas atuam sobre a esfera. A for\c ca peso, m{\bf g}, 
atuante na dire\c c\~ao vertical para baixo, e a for\c ca {\bf N} 
de rea\c c\~ao da superf\'{\i}cie do cone, atuante na dire\c c\~ao 
normal \`a superfi\'{\i}cie quando se despreza o atrito. 
Haver\'a um 
valor prop\'{\i}cio da velocidade inicial $v$ para o qual a esfera 
efetuar\'a um movimento circular de raio $r=s\cos{\theta}$ com certa 
velocidade angular $\omega$ em torno de um eixo vertical que passa 
por $O$. Isso ocorrer\'a quando a resultante das duas for\c cas que 
atuam sobre a esfera produzirem uma resultante centr\'{\i}peta 
situada no plano do movimento circular.
\newline
As equa\c c\~oes do movimento da esfera s\~ao obtidas decompondo-se a 
for\c ca peso e a for\c ca de rea\c c\~ao nas dire\c 
c\~oes vertical e horizontal. Disso vem
\be\label{1um}
N \cos{\theta}=mg
\ee
para as componentes verticais das for\c cas que atuam na esfera. As 
componentes horizontais das for\c cas que atuam na esfera produzem a 
acelera\c c\~ao centr\'{\i}peta do movimento. Neste caso se tem que
\be\label{2ois}
N\sin{\theta}=m\frac{v^{2}}{r}=m\omega^{2}s \cos{\theta}.
\ee
Portanto, ao se eliminar $N$ das equa\c c\~oes (\ref{1um}) e 
(\ref{2ois}), obt\'em-se a seguinte express\~ao para a velocidade 
angular da esfera
\be\label{omega1}
\omega^{2}=\frac{g\sin{\theta}}{s \cos^{2}{\theta}},
\ee
e, da rela\c c\~ao $v=\omega r$, obt\'em-se a velocidade 
que se deve imprimir \`a esfera para que o movimento circular
desejado seja poss\'{\i}vel. 
\newline
\noindent
\noindent
No problema que acabamos de discutir, a quest\~ao do plano do 
movimento circular \'e imposta na formula\c c\~ao do problema,
e isso nos leva \`a decomposi\c c\~ao das for\c cas nas 
dire\c c\~oes vertical e horizontal praticamente sem nenhuma hesita\c 
c\~ao. A quest\~ao do plano do movimento circular parece admitir mais 
de uma interpreta\c c\~ao no problema ilustrado na {\sc figura} 
\ref{con2} (a). Aqui uma esfera de massa $m$ move-se em um trilho 
circular de raio $s$ e centro $O$ sobre um plano inclinado de um 
\^angulo $\theta$ com a horizontal. Pode-se perguntar qual deve ser a 
velocidade da esfera na parte mais alta do trilho para que ela possa 
descrever a circunfer\^encia nessa parte mais alta do plano inclinado.
\newline
\'E claro que a esfera vai descrever uma c\'{\i}rculo no plano inclinado, 
j\'a que o trilho est\'a nesse plano. Por isso, parece ser mais natural 
decompor as for\c cas que atuam sobre a esfera como mostrado 
na {\sc figura} \ref{con2} (b) no momento em que a esfera est\'a na 
parte mais alta do plano inclinado. Essa decomposi\c c\~ao \'e
condizente com a decomposi\c c\~ao usual das for\c cas em um movimento 
circular genu\'{\i}no, de raio $s$ e centro $O$, no plano horizontal,  
como discutido no problema anterior, com a \'unica diferen\c ca de 
que o plano movimento circular agora \'e inclinado. 
\newline
Perceba tamb\'em que neste caso o vetor velocidade angular da esfera 
ser\'a certo vetor ${\bf \Omega}$ perpendicular ao plano inclinado,
e portanto perpendicular ao raio do movimento circular, exatamente 
como o era na {\sc figura} \ref{conn}.
\newline
N\~ao ser\'a poss\'{\i}vel, no entanto,  -- e talvez seja mesmo 
v\'alido --, considerar que o plano do movimento circular \'e 
horizontal tamb\'em na situa\c c\~ao descrita na {\sc figura} 
\ref{con2}? Pode-se ir al\'em e considerar o movimento circular
da esfera em qualquer plano que passe pela posi\c c\~ao da 
esfera?
\newline
Para responder a essas indaga\c c\~oes, vamos primeiramente analisar a
situa\c c\~ao descrita na {\sc figura} \ref{con2} (b). Como antes, as 
duas for\c cas que atuam na esfera s\~ao seu peso e a for\c ca de rea\c 
c\~ao normal ao plano. Decompondo-se as for\c cas peso e de rea\c c\~ao 
nas dire\c c\~oes perpendicular e paralela ao plano, obt\'em-se
\be\label{33}
N=mg\cos{\theta}
\ee
para as componentes perpendiculares, e, para as componentes paralelas,
obt\'em-se a resultante centr\'{\i}peta:
\be\label{44}
mg\sin{\theta}=m\frac{v^{2}}{s}=m\Omega^{2}s,
\ee
sendo $v$ a velocidade da esfera no ponto mais alto do plano inclinado, 
e $s$ o raio do movimento circular. Novamente eliminando $N$ entre
as equa\c c\~oes (\ref{33}) e (\ref{44}) vem
\be\label{omega2}
\Omega^{2}=\frac{g\sin{\theta}}{s}.
\ee
A quest\~ao que se nos apresenta \'e se se pode considerar
que a resultante das for\c cas que atuam na esfera na situa\c c\~ao 
descrita na {\sc figura} \ref{con2} est\'a na dire\c c\~ao horizontal, 
em vez de na dire\c c\~ao paralela ao plano inclinado. Isso \'e o 
mesmo que perguntar: h\'a alguma rela\c c\~ao entre as f\'ormulas 
(\ref{omega1}) e (\ref{omega2})? A resposta \'e que podemos consider 
que o movimento circular ocorra em qualquer plano que passe pela esfera, 
contanto que usemos a correta decomposi\c c\~ao dos vetores envolvidos no 
movimento. Assim, o vetor velocidade angular ${\bf \Omega}$ do movimento 
circular na {\sc figura} \ref{con2} est\'a de fato relacionado com um 
vetor velocidade angular ${\bf w}$ na dire\c c\~ao vertical, como se 
o movimento circular da esfera ocorresse no plano horizontal. Isso vem
da equa\c c\~ao 
\[
\Omega=w \cos{\theta},
\]
a qual \'e obtida por simples decomposi\c c\~ao vetorial como
mostrado na {\sc figura} \ref{con2}. Substituindo essa express\~ao 
em (\ref{omega2}), obtemos 
\[
w^{2}=\frac{g\sin{\theta}}{s \cos^{2}{\theta}}.
\]
Essa equa\c c\~ao tem a mesma forma que a equa\c c\~ao (\ref{omega1}),
preservando ainda a mesma rela\c c\~ao entre o raio $s$ do 
movimento circular e o vetor velocidade angular.
\noindent
\subsection{Exemplos}
\begin{figure}
\begin{center}
\resizebox{!}{9 truecm}{\includegraphics{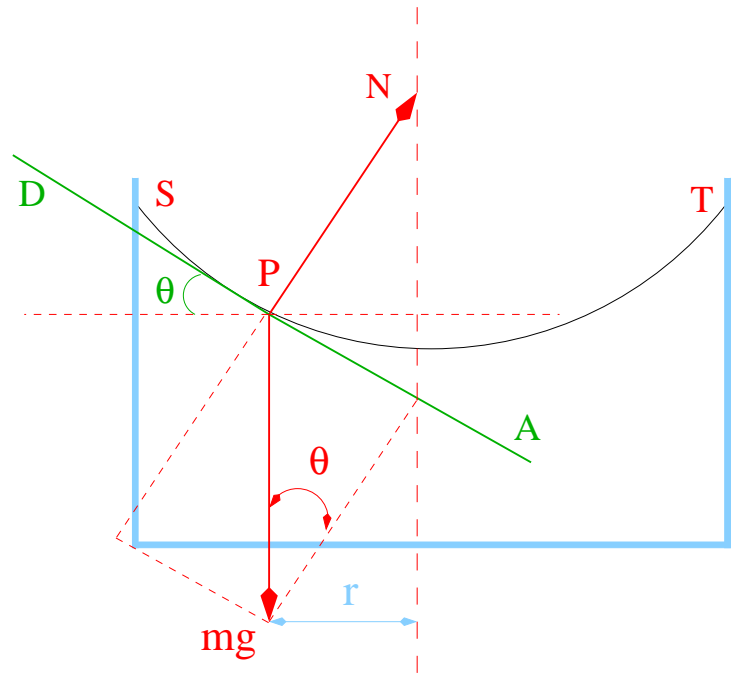}}
\vskip .1 cm
\parbox{14truecm}{\caption{\baselineskip=16 pt\small\label{eliq}
\noindent
O perfil (SPT) da superf\'{\i}cie de um l\'{\i}quido em um 
vaso cil\'{\i}ndrico que gira em torno de um eixo vertical
com velocidade angular constante.}}
\end{center}
\end{figure}

\noindent
{\bf 1)} Seja ${\bf \omega}$ o vetor velocidade angular de 
um corpo em movimento circular. Sejam $O\xi$ e $O\eta$ dois
eixos quaisquer no plano do vetor velocidade angular, e 
suponha que o vetor ${\bf \omega}$ forme \^angulos $\alpha$ 
e $\beta$, respectivamente, com os eixos dados. Demonstre que 
as componentes de ${\bf \omega}$ nas dire\c c\~oes dos eixos 
$O\xi$ e $O\eta$ s\~ao dadas pelas f\'ormulas
\[
\omega_{\xi}=\frac{\omega \cdot \sin\beta}{\sin(\alpha+\beta)} 
\;\;\;\;\;\;\;\;\;\;\;\;\; {\rm and} \;\;\;\;\;\;\;\;\;\;\;\;\; 
\omega_{\eta}=\frac{\omega \cdot \sin\alpha}{\sin(\alpha+\beta)}.
\]
O que se obt\'em quando os \^angulos $\alpha$ e $\beta$ s\~ao tais
que $\alpha+\beta=\frac{\pi}{2}$?
\vskip 1 cm
\noindent
{\bf 2)} Seja um l\'{\i}quido contido em um vaso cil\'{\i}ndrico
dotado de um movimento de rota\c c\~ao com velocidade angular $\Omega$
em torno de seu eixo vertical de simetria. A superf\'{\i}cie livre do 
l\'{\i}quido se curva formando um menisco. A {\sc figura} \ref{eliq} 
mostra uma part\'{\i}cula $P$ de massa $m$ na superf\'{\i}cie do 
l\'{\i}quido. A part\'{\i}cula descreve um movimento circular 
de raio $r$ com velocidade angular $\Omega$. Seja $DA$ a
tangente \`a superf\'{\i}cie do l\'{\i}quido no ponto $P$.  
A tangente $DA$ forma um \^angulo $\theta$ com a horizontal, como
mostrado na figura. Duas for\c cas atuam sobre $P$, a saber: o peso m{\bf 
g} da part\'{\i}cula que atua verticalmente para baixo, e a for\c ca 
normal {\bf N} que atua perpendicular \`a dire\c c\~ao de $DA$. A 
for\c ca normal \'e a resultante das for\c cas que o restante do 
l\'{\i}quido exerce sobre $P$. A soma das for\c cas peso
e normal \'e uma resultante centr\'{\i}peta que atua horizontalmente 
na dire\c c\~ao do eixo de rota\c c\~ao do vaso que cont\'em o 
l\'{\i}quido. 
\vskip .1 cm
\noindent
{\bf (a)} Mostre que no caso de equil\'{\i}brio, o \^angulo $\theta$ \'e 
tal que\cite{P}
\[
{\rm tg}{\theta}=\frac{\Omega^{2}r}{g}.
\]
{\bf (b)} Portanto, para uma dada velocidade angular $\Omega$, a 
tangente de $\theta$ \'e diretamente proporcional ao raio do movimento 
circular. Mostre que a par\'abola \'e a curva para a qual o
\^angulo de inclina\c c\~ao da tangente \`a curva \'e diretamente
proporcional \`a dist\^ancia do eixo da par\'abola. Use esse fato
para concluir que a condi\c c\~ao no \^angulo $\theta$ obtida acima
implica que o perfil da superf\'{\i}cie l\'{\i}quida mostrado na {\sc 
figura} \ref{eliq} \'e uma se\c c\~ao de um parabol\'oide de revolu\c 
c\~ao.
\vskip .8 cm
\noindent
{\bf 3)} Um ponto material de massa $m$ \'e livre para se mover 
sobre uma circunfer\^encia de raio $R$ mantida em uma plano vertical. 
O ponto mant\'em-se im\'ovel em rela\c c\~ao \`a circunfer\^encia 
quando esta \'e dotada de movimento de rota\c c\~ao com velocidade
angular constante $\omega$ em torno de seu di\^ametro vertical.
\vskip .1 cm
\noindent
{\bf (a)} Mostre que, nas condi\c c\~oes do problema, a posi\c c\~ao
de equil\'{\i}brio do ponto sobre a circunfer\^encia, medida
pelo \^angulo $\theta$ que o raio da circunfer\^encia forma com a 
vertical naquela posi\c c\~ao do ponto, \'e tal que
\[
\cos{\theta}=\frac{g}{\omega^{2}R}.
\]
{\bf (b)} A posi\c c\~ao de equil\'{\i}brio encontrada acima n\~ao 
\'e \'unica. Quantas posi\c c\~oes de equil\'{\i}brio existem 
quando $g \leq \omega^{2}R$? 
\vskip .8 cm
\noindent
{\bf 4)} Uma barra $AB$ de comprimento $l$ move-se em um plano de tal 
maneira que as velocidades de suas extremidades $A$ e $B$ formam 
com a barra \^angulos $\alpha$ e $\beta$, respectivamente. Prove 
que a velocidade angular da barra \'e dada por
\[
\frac{u\sin{\left (\alpha-\beta \right )}}{l\cos{\beta}},
\]
sendo $u$ a velocidade da extremidade $A$.
\vskip .5 cm
\noindent
\begin{center}
{\bf Refer\^encias}
\end{center}

\vfill
\hrule width2truein \smallskip {\baselineskip=10pt \noindent
{\small Uma estudante come\c cou a estudar para uma prova de 
F\'{\i}sica entre as 3 e as 4 horas da madrugada quando os 
ponteiros de seu rel\'ogio coincidiam. A estudante interrompeu 
seus estudos, vencida pelo sono, quando os ponteiros de seu 
rel\'ogio novamente coincidiam entre as 5 e as 6 horas
da manh\~a. Quanto tempo estudou a estudante?}
\par
}

\end{document}